\documentstyle[12pt]{article}
\begin{document}
\baselineskip = 20pt

\begin{flushright}
{IF-UFRJ/99}
\end{flushright}
\vspace*{5mm}

\begin{center}

{ HAWKING RADIATION IN THE DILATON GRAVITY\\ WITH A NON-MINIMALLY COUPLED\\ SCALAR FIELD} 

\vspace*{1cm}

{M. ALVES$^{\star}$ \\
\vspace{0.3cm}
Instituto de Fisica - UFRJ\\
Rio de Janeiro-RJ \\
Brazil}\\
\vspace*{1cm}
{ABSTRACT}
\end{center}
\vspace*{2mm}

\noindent
We discuss the two-dimensional dilaton gravity with a scalar field as 
the source matter where the coupling with gravity is given, besides the minimal one, through an external field. This coupling generalizes the conformal anomaly in the same way as those found in recent literature, but with a diferent motivation. The modification to the Hawking radiation is calculated explicity and show an additional term that introduces a dependence on the (effective) mass of the black-hole.

\bigskip
\vfill
\par
\bigskip
\bigskip

\par
\bigskip
\bigskip
\noindent $^\star${e-mail: MSALVES@if.ufrj.br}

\pagebreak
\noindent
{\bf 1 INTRODUCTION}
\par It is widely recognized that two-dimensional models of gravity can give us a better understanding of the gravitational quantum effects. These models, derived either from a string motivated effective action [1] or from some low-dimensional version of the Einstein equations [2], have a rich structure in spite of their relative simplicity. Gravitational collapse, black holes and quantum effects are examples of subjects whose description is rather complicated in four-dimensional gravity while their lower-dimensional versions turn out to be more treatable, sometimes completely solved. This is the case of the seminal work of Callan, Giddings, Harvey and Strominger (CGHS)[1], where black-hole solutions are found and  analysed semi-classically, giving us a two dimensional version of the Hawking effect.

\par In the CGHS model, the starting point is the four-dimensional Einstein-Hilbert action in the spherically symmetric metric, where the Schwarzchild one is the simplest case. Then, the assumption of dependence in two variables for all the fields is done and,  with a suitable form of the metric, it is possible to integrate the angular dependence. The final result is a two-dimensional action with a new field, besides the gravitational one, called the dilaton, that can be taken as the relic of the integrated coordinates. This is the two-dimensional dilaton gravity.
Some improvements have been done [3] to circumvent difficulties that  arise from its quantum version, but leaving the initial purpose unchanged.

\par Black-hole solutions are found to be formed from
non-singular initial conditions, namely a source of scalar matter
coupled minimally with the two-dimensional gravity sector (the terminology
will be clarified later). There is also a linear vacuum region,
which turns to be relevant to these models. 

 Since these results are classical, the next step is the search for quantum effects: it is well known that whereas 
classical black-holes radiate nothing the semi-classical ones render a
thermal radiation by a proccess  called Hawking-Beckestein effect. 
 An important feature of the semi-classical version for this effect is the conformal anomaly, since the Hawking radiation can be derived from this quantity. Recently, a series of works[10] gives atention to the generalization of the conformal anomaly to the CGHS model, with diferent coupling between the scalar field and gravitation. 

By semi-classical version of this theory we mean the scalar field quantized in the curved, classical, background. However , as pointed out before [6,7,8], we must be careful with the field variable to be considered when we use the Fujikawa method.
In the present case we make a redefinition of the scalar field variable that, in the two dimensional case, results in a more general expression for the trace anomaly. We discuss here how this redefinition generates the same expression to the generalized anomaly found in [10] in a simple and direct way. This is one of the results of this work. 

 On other hand, since the anomaly is used to calculate the Hawking radiation in the two dimensional case, we can expect new contributions to this quantity from the new terms of the conformal anomaly.  In the CGHS model, these new terms give a generalization to the expression of the Hawking radiation  that depend of the effective mass of the black-hole, absent in the original calculation. 

\par This article is organized as follows: in the next section we discuss the motivation for the definition of a new field variable, the modification in the original action and the resulting expression for the trace anomaly in a direct way.  After that, we present without details the main results of the CGHS model concerning the black hole radiation to fix notation, then, following the same steps as in [1], we calculate the new expression for the Hawking radiation derived from the modified anomaly. Discussions and final remarks are in the conclusion.

\pagebreak
\bigskip
\bigskip
\bigskip
\noindent
{\bf 2  THE 2d DILATON GRAVITY WITH NON-MINIMAL COUPLING }

The semi-classical quantization via functional method requires the integration
over fields (the scalar one in this particular case). On the other hand, we
are interested in theories that have the full quantized version free of 
anomalies. This paradigm can be worked out via the BRST analysis of the
theory by using the Fujikawa's technique [6,7]. In this framework, the quantized theory is anomaly-free provided the functional measure is BRST-invariant. It follows that this invariance requires a redefinition of the field variables, the so-called gravitational dressing, and we must consider these new fields as the variables of the model that we are studying. We stress that without this modification the conservation of the quantized energy-momentum tensor (EMT) is not verified [7]. Another remarkable fact is that the trace anomaly would be null (only in the 2d case). 

 For the present case, 2d scalar field $f$, this redefinition means    

\begin{equation}
f \rightarrow  \tilde f = (-g)^{({1\over 4})} f,
\end{equation}
where $g = det g_{\mu\nu}$.

The original action can be rewriten in terms of this new variable:

\begin{equation}
S[f,g_{\mu\nu}]={1\over{2\pi}}\int d^{2}x \sqrt{-g}\, \nabla_{\mu}f\nabla^{\mu}f \rightarrow
S[\tilde f,g_{\mu\nu}]={1\over{2\pi}}\int d^{2}x \, \nabla_{\mu}\tilde f\nabla^{\mu}\tilde f
 \end{equation}

It is straightforward to see that the action for the new variables is not conformal invariant, since by a transformation as 

\begin{equation}
 g_{\mu\nu}^{'} = e^{2\alpha}g_{\mu\nu}
\end{equation}

\noindent the field $\tilde f$ transform as

\begin{equation}
\tilde f \rightarrow \tilde f^{'}= e^{\alpha}\tilde f.
\end{equation}
Now, let us use the conformal gauge

\begin{equation}
g_{\mu\nu} = e^{2\rho}\eta_{\mu\nu}\;\;\;\;\;\;\; (-g)^{({1\over 4})}=e^{\rho}
\end{equation}
and write 

\begin{equation}
\tilde f^{'} = (e^{\rho} f)^{'} = e^{\alpha}e^{\rho} f
\end{equation}
\noindent so, in this gauge, a conformal transformation on $f^{'}$ is equivalent to make
\begin{equation}
\rho \rightarrow \rho^{'} = \rho + \alpha
\end{equation}

Using these relations, we can  define a non-minimal coupling to the field variable through a gauge type field, namely

\begin{equation}
\tilde\nabla_{\mu}\tilde f = (\nabla_{\mu} - A_{\mu})\tilde f
\end{equation}

\noindent where

\begin{equation}
A_{\mu} = \nabla_{\mu}\rho
\end{equation}
tranforms as

\begin{equation}
 A_{\mu}^{'} =A_{\mu} + \nabla_{\mu}\alpha
\end{equation}

With the definition (8), we can write a conformally invariant action to the fields $\tilde f$ and $g_{\mu\nu}$:

\begin{equation}
S[\tilde f,g_{\mu\nu}]={1\over{2\pi}}\int d^{2}x \,\,\tilde\nabla^{\mu}\tilde f \tilde\nabla_{\mu}\tilde f
\end{equation}

\noindent or 
\begin{equation}
S[\tilde f,g_{\mu\nu}]=-{1\over{2\pi}}\int d^{2}x \,\,\tilde f(\nabla^{\mu}\nabla_{\mu}-A^{\mu}A_{\mu} + \nabla^{\mu}A_{\mu})    \tilde f
\end{equation}

\par This modification leads to a generalization of the value of
trace anomaly  which is easily calculated, since the field $A_{\mu}$ is not quantized and can be considered as an external field. The resulting anomaly is [8]: 

\begin{equation}
 R \rightarrow R_{(generalized)} = R + \beta ( A^{\mu}A_{\mu} - \nabla^{\mu}A_{\mu} ) 
\end{equation}

\noindent  Here, the parameter $\beta$ shows us the presence of non-minimal coupling ($\beta=1$) or its absence ($\beta=0$).

Antecipating a result from the next section, to wit the conformal factor $\rho$ equal to the dilaton field $\phi$, we have:
\begin{equation}
 R_{(generalized)} = R + \beta ( \nabla^{\mu}\phi \nabla_{\mu}\phi + \nabla_{\mu}\nabla^{\mu}\phi ) 
\end{equation}

\par There are many recent works [10] dealing with this general value
for the 2d trace anomaly, some of them with others but related motivation , rendering different  numerical values for the extra terms in (14). This does not change our analysis.

\pagebreak
\bigskip
\bigskip
\bigskip
\noindent
{\bf 3  THE CGHS MODEL FOR THE 2d DILATON GRAVITY AND THE HAWKING RADIATION}

\par
Intending to compare results, we present in this section 
the CGHS model for the two dimensional gravity and, specifically, the expression
of the radiation of the black hole, derived from the semi-classical version following the same steps as in [1]. 
\par The starting point is the action
\begin{equation}
S={1\over{2\pi}}\int d^{2}x \sqrt{-g}\, e^{-2\phi} \biggl\{ R + 
4(\nabla\phi)^{2} +  4\lambda^{2} \biggr\}.
\end{equation}

Here, $R$ is the bidimensional scalar curvature and $\lambda$ is to be considered as a cosmological constant.
As mentioned before, the dilaton field $\phi$, that in two dimensional space time is a scalar, came from the angular part of the
original 4d metric , so that this action must be considered as the pure gravitational part. Stressing this affirmative is the possibility to make the dilaton equal to the gravitational field, at least classically.

\par Using the light-cone coordinates and the conformal gauge, the metric becomes

\begin{equation}
g_{+-} = -{1\over 2} e^{2\rho}\, , \,\,\,\, g_{++} = g_{--} = 0.
\end{equation}

\par This two dimensional model shows a black hole type solution when scalar fields $\varphi$ are considered as matter source in $x^{+}=x^{+}_{0}$, traveling in the $x^{-}$ direction, 
  with the intensity proportional to the constant $a$.

The solutions for the resulting equations are:

\begin{equation}
e^{-2\rho} = e^{-2\phi} = {M\over{\lambda}}-{\lambda}^2 x^{-}x^{+}
\end{equation}
\noindent for $x^{+} > x^{+}_{0}$  and for $x^{+} < x^{+}_{0}$ we have the vacuum.  $M=ax_{0}^{+}\lambda$ is identified with the mass of the hole. 
Note that $\rho = \phi$ is a consequence of the calculation and it will be used later.

 Up to this point, all the results are classical and to obtain information about the Hawking radiation we must consider quantum effects. This can be done through
the relation between the trace anomaly and the components of the vacuum expectation value (VEV) of the energy momentum  tensor (EMT) [9] . In two dimensions,
the VEV for the massless field is given by

\begin{equation}
\langle T^{\mu}_{\mu} \rangle = {N\over 24} R 
\end{equation} 
  
\noindent where $R$ is the scalar curvature and $N$ is related with the number of the fields in the model. Of course, the RHS of this equation is only due to the quantum corrections since the classical expression for the trace of the EMT is null for zero mass fields. Up to numerical factors, the expression to this anomaly is the same to all kinds of fields.

The conservation of the EMT must be imposed, so 

\begin{equation}
\nabla^\nu\langle T_{\mu\nu}\rangle = 0.
\end{equation}

The solutions of these equations  are, in the conformal gauge,

\begin{equation}
T_{--} = \partial_{-}\rho\partial_{-}\rho - \partial^{2}_{+}\rho
+ t_{-}(x^{+}) 
\end{equation}

 and

\begin{equation}
T_{++} = \partial_{+}\rho\partial_{+}\rho - \partial^{2}_{-}\rho
+ t_{+}(x^{-})
\end{equation}

The limits of (20) and (21) at the assimptotical regions , $I^{+}$ and $I^{-}$, give us the value of $t_{+}$ and $t_{-}$.
In order to compare with the literature, we use the same coordinates as in [CGHS]:

\begin{equation}
x^{+} =  {1\over\lambda} e^{\lambda y^{+}}\;\;\;\mbox{and}\;\;\;
x^{-} =  -{1\over\lambda} e^{\lambda y^{-}} + {a\over \lambda^{2}}.  
\end{equation}

These definitions result in a new metric but conformally related
with the flat one like the former or, in others words, they preserve
the conformal gauge. This new metric is obtained by writing (17) in terms of the variables $y^{+}$ and $y^{-}$ giving the conformal factor as

\begin{equation}
e^{2\rho}=
\begin{array}{l}

(1+{a\over \lambda}  e^{\lambda y^{-}})^{-1}\;\;\;\mbox{for}\;\;\;y^{+}<y_{0}^{+}\\ \\ \\

(1+{a\over \lambda}  e^{\lambda (y^{-}-y^{+}+y^{+}_{0})})^{-1} 
\;\;\;\mbox{for}\;\;\;y^{+}>y_{0}^{+},
\end{array}
\end{equation}

\noindent where $\lambda x^{+}_{0} = e^{\lambda y^{+}_{0}}$.

The requeriment  that these expressions vanish in the vacuun ( $x^{+}<
x^{+}_{0}$) gives us the values of $t_{+}$ and $t_{-}$ and can be
calculated at the limits  $e^{-\lambda y^{-}}\rightarrow \infty$ and $e^{-\lambda y^{-}}\rightarrow -{a\over\lambda}$. This condition applied in (20), give us

\begin{equation}
 t_{-}={-\lambda^{2}\over 4}(1-(1+{a\over \lambda}e^{\lambda y^{-}})^{-2}).
\end{equation}

The value for the $T_{--}$ component at $y^{+}\rightarrow \infty $ ($x^{+}\rightarrow \infty $) is the flux across the future null infinity and, taken this limit again in (20), we see that the remaining term  is just 
$t_{-}$.
 
\par The Hawking radiation is the value of the $T_{--}$ near the horizon ( at $y^{-}\rightarrow \infty$, it vanishes):

\begin{equation}
x^{-}\rightarrow-{a\over \lambda^{2}} \;\;\;\;\mbox{or}\;\;\;\;
e^{-\lambda y^{-}}\rightarrow 0 ,
\end{equation}

Using (24) and (20), the final expression is:
\begin{equation}
T_{--}^{horizon}= {\lambda^{2}\over 4}.
\end{equation}

\par At this point, it is worth mentionning that the absence of the mass in (26) is peculiar of the CGHS model.

Now, let us consider the modifications in the expression to the anomaly and show how this result affects the expression for the Hawking radiation.
Our strategy will be to follow the same steps as those showed above, using the expression (13) in (18) and (19).

\par With our choice to the gauge, the trace anomaly is now given by 

\begin{equation}
R = e^{-2\rho}(8\partial_{+}\partial_{-}\rho -6\beta (16\partial_{+}\partial_{-}\rho + 64\partial_{+}\rho\partial_{-}\rho)).
\end{equation}

The non-zero components of relation (18) turns to be

\begin{equation}
T_{+-} = \alpha\partial_{+}\partial_{-}\rho +
  \beta\partial_{+}\rho\partial_{-}\rho,
\end{equation}

\noindent where the constants $\alpha$ and $\beta$ were redefined in terms of those ones of (13). In this way, it is simple to see that the new contribution comes from the second term in RHS of (28).   

\par Following the same steps as before, namely using (28) in the (18) and (19), we have , e.g, for the $T_{--}$ component

\begin{equation}
T_{--}\; = \;
T_{--}^{\beta=0}\; + \;
\beta [{1\over 2}(\partial_{-}\rho)^{2} + \rho\partial_{-}^{2}\rho - 2\rho(\partial_{-}\rho)^{2}] + t(y^{-})
\end{equation}

\noindent where $T_{--}^{\beta=0}$ is given by (20) or (21), depending on which portion of the space time we are considering: at the $y^{+}<y^{+}_{0}$ region, there is no extra contribution, since the new terms do not depend on $y^{+}$. Consequently $t(y^{-})$ has the same value calculated previously. 

\noindent 

\par As before, the expression for  $t(y^{-})$ at  $y^{+}>y^{+}_{0}$is obtained by applying the assimptotical limit at the vacuum region, $y^{-}\rightarrow 0$:

\begin{equation}
t(y^{-})\; = \; t(y^{-})^{\beta =0} \; + \; \beta {\lambda\over 4}({1\over 2}+{1\over 2}e^{4\rho} - 2\rho),
\end{equation} 
with  $t(y^{-})^{\beta =0}$ given by(26) and $\rho$ by (23). 

\par Finally, taking the appropriated limits, we arrive at the desired expression,

\begin{equation}
T_{--}^{horizon} \; = \; {\lambda^{2}\over 4} [ \alpha + \beta ln({M\over a}) ]
\end{equation}

\noindent where the constants $\alpha$ and  $\beta$ were redefined again for simplicity.

The modification of the expression of the Hawking radiation (31) was already expected, since the expression to the anomaly was modified. The dependence with the effective mass arises due the nonlinearity of the extra term in (27) and is a direct consequence of the new couplings introduced before. This is another result of this work.

\pagebreak

\bigskip
\bigskip
\bigskip
\noindent
{\bf 4  CONCLUSIONS AND FINAL REMARKS}

In this paper, we use the non-minimal coupling between the scalar field and  2d gravitation that  gives rise to a generalization to the trace anomaly. The non-minimal coupling includes a type-gauge field , given in terms of the gravitation (or the dilaton in the CGHS model ) field. The motivation of the introduction of this extra coupling is the requirement of the conformal invariance of the action for the redefined field that, differently to the original, is not conformally invariant.

 The semiclassical quantization allows us to consider this gauge field as an external one, rendering the calculation of the anomaly very simple: we just need to add the new terms that appear in the equation of motion derived from (12).

The expression for the auxiliary field  is due to the fact that in two dimensions we allways can use the conformal gauge. In higher dimensions, this choice would be very restrictive but, in this case, it is not nescessary to redefine the matter field to satisfy (19) and the definition of the conformal gauge field is straightfoward [13] . Actually, in the 4d case, (19) is used to fix some of numerical values of the EMT [5].

\par In two dimensional space-time the calculation of the Hawking radiation is easily obtained via its relation with the trace anomaly [9]. The calculations using the CGHS model yield a expression for black hole radiation that do not depend on the mass of the hole. On the contrary, in the Schwarzchild black hole, there is such dependence and has important consequences on the behaviour of these objects[9].

When we use the generalized conformal anomaly the resulting expression to the Hawking radiation has a dependence with the mass. The modification in the Hawking radiation was expected because the relations (13) and (18) and we can expect a more general behaviour for these structures. We mention also that, in the CGHS model, the dependence of the radiation with the temperature is assured by the relation with the $\lambda $ parameter, since the temperature in this case is given by [12]:

\begin{equation}
T = {\lambda\over 2\pi} 
\end{equation}

\noindent and it is found to be the same as in the four dimensional case, to wit, $ T_{--}\sim T^{2}$  
(in this case, $ T\sim M  \;\; $  and $\;\; T_{--}\sim M^{2}$). However, the modification in (31)  breaks the relation between radiation and temperature so this result  must be taken as a higher order term and not as the complete expression.

Finally, it must be interesting to study others  models  using the non-minimal coupling to see what are the consequences of this choice. Works in this direction are in progress.

\bigskip
\bigskip
\bigskip
\pagebreak
\noindent
{\bf ACKNOWLEDGEMENTS}

The author is grateful to Prof. Carlos Farina for reading the manuscript and useful comments. This work was partially supported by Funda\c c\~ao Universit\'aria Jos\'e Bonif\'acio, FUJB.

\bigskip
\noindent
{\bf REFERENCES}

\noindent [1] C.G. Callan, S.B. Giddings and J.A. Strominger, 
Phys. Rev. D 45 (1992)
R1005; J.A. Strominger, in Les Houches Lectures on Black Holes (1994), 
hep-th/9501071.

\noindent [2] R. Mann, A. Shiekm and L. Tarasov, Nucl. Phys. B341 (1992) 134; 
R. Jackiw, in Quantum Theory of Gravity, ed. S.Christensen (Adam Hilger,Bristol,
1984), p.403; C. Teitelboim, ibid, p. 327.

\noindent [3] J. Russo, L. Susskind and L. Thorlacius, 
Phys. Rev. D 45 (1992) 3444;
47 (1993) 533

\noindent [4] J. Maharana and J.H. Schwarz, Nucl. Phys. B390 (1993) 3; 
J.Scherk and J.H.
Schwarz, Nucl. Phys. B153 (1979) 61; J. Maharana, 
Phys. Rev. Lett. 75,2 (1995) 205.

\noindent [5] N.D. Birrel and P.C. Davies, in 
Quantum Fields in Curved Spacetime 
(Cambridge University Press, Cambridge, 1984)

\noindent [6] K. Fujikawa in Quantum Gravity and Cosmology, ed. H.Sato
and T.Inami (Singapore: World scientific); Phys. Rev. D 25 (1982) 2584.

\noindent [7] K. Fujikawa, U. Lindstrom, N.K. Rocek and P.van Nieuwenhuizen, 
Phys. Rev. D 37 (1988) 391. 

\noindent [8] M.Alves and C.Farina, Class.Quantum Grav. 9 (1992)1841; 
M.Alves, Class.Quantum Grav. 13 (1996) 171.

\noindent [9] S.M.Christensen and S.A.Fulling, Phis.Rev.D 15 (1977) 2088.
 
\noindent [10] S.Hawking and R.Boussos, hep-th/9705236; J.S.Dowker, hep-th/9802029;
 S.Ichinoise and S.Odintsov, hep-th/9802043.

\noindent [11] S.Hawking, Commun.math.Phys.43,199(1975);G.Gibbons and S.Hawking,
Phys.Rev.D 15(1976)2738.

\noindent [12] See, for example, A.Gosh, hep-th/9604056 and references therein.

\noindent [13] M.Alves and J.Barcelos-Neto, Class.Quantum Grav. 5(1988)377.
\end{document}